\documentclass[11pt]{article}
\usepackage{DGfest,epsfig}
\usepackage{times}

\def\be{\begin{equation}}
\def\ee{\end{equation}}
\def\bea{\begin{eqnarray}}
\def\eea{\end{eqnarray}}

\newcommand{\eq}{\begin{equation}}
\newcommand{\beq}{\begin{equation}}
\newcommand{\eqx}{\end{equation}}
\newcommand{\eeq}{\end{equation}}
\newcommand{\eqn}{\begin{eqnarray}}
\newcommand{\eqnx}{\end{eqnarray}}
\newcommand{\ad}{{a^{\dagger}}}
\newcommand{\fd}{f^{\dagger}}

\newcommand{\Qd}{{Q^{\dagger}}}

\newcommand{\nn}{\nonumber}
 \newcommand{\None}{${\cal N} =1~$}

\newcommand {\eqref} [1] {(\ref {#1})}
\newcommand {\slsh} [1] {\not{\hbox{\kern-2pt${#1}$}}}

\begin{document}
\baselineskip 11.5pt

\begin{flushright}
\vspace{15mm}
CERN-PH-TH/2006-035\\
TPJU-1/2006
\end{flushright}

\title{LARGE $N$,  SUPERSYMMETRY \dots  AND QCD }

\author{  G. Veneziano }

\address{Theory Division, CERN, CH-1211 Geneva 23, Switzerland \\
and \\
Coll\`ege de France, 11 place Marcelin Berthelot, 75005 Paris, France}

\author{  J. Wosiek  }

\address{M. Smoluchowski Institute of Physics, Jagellonian University\\
Reymonta 4, 30-059 Krak\'{o}w, Poland }

\maketitle\abstracts{
 This paper consists of two (still only vaguely) related parts: in the first,
 we  briefly review work done in the past three  years on the ``planar equivalence"
 between a class of non-supersymmetric  theories (including limiting cases of QCD)
 and their corresponding supersymmetric ``parents";  in the second, we present  details of   a new
 formulation of  planar  quantum mechanics and  illustrate its effectiveness in an intriguing
 supersymmetric example.}

\section{Introduction}
Lattice Gauge Theory (LGT)
--an approach where Adriano has distinguished himself with many important achievements--  is the main tool available at present for extracting  quantitative results from QCD. This approach
 has established unambiguously some key non-perturbative features of QCD, such as
confinement, chiral symmetry breaking, the absence of a $U(1)$-Nambu--Goldstone (NG) boson, and more.
In spite of these uncontested successes, LGT remains limited, in its predictive power, by  its
range of applicability (in terms of number and kind of external particles and of the accessible kinematical regions) and
by the fact that most of its results, being  mere numerical computer outputs, often offer  limited theoretical insight.

 Few  techniques are
available today for the {\it analytic} study of non-perturbative  QCD.
Among these,  the method of  effective chiral lagrangians, based on the general concepts of spontaneous symmetry breaking (SSB) and (pseudo) Nambu--Goldstone bosons is one of the most widely used. Even though successful, it  is also  limited in scope to  the spectrum and interactions of light NG bosons. At the other extreme, heavy quark effective theories have been of great value for describing the behaviour of hadrons carrying charm and beauty.

A very different kind of analytic tools is given, finally,   by large-$N$ expansions of various kinds, an approach that is a priori of  much wider applicability.
Formulated more than three decades ago \cite{'tH} \cite{TE}, large-$N$ limits still remain elusive while attracting
the attention of many theorists \cite{BW}. The relation between the large-$N$ classification of diagrams
and loop expansions in string theory only enhances the interest in this approach.
Unfortunately, with the exception of some quantitative results for the $\eta'$ mass and interactions \cite{WV},  large-$N$ predictions have been mostly qualitative in nature.

The situation concerning quantitative analytic results in strongly coupled gauge theories improves considerably if one is willing to consider supersymmetric variants --or extensions-- of  QCD.
Predictivity is increased dramatically with the number ${\cal N}$ of supersymmetries:
while for
${\cal N} =1$ this is basically  limited to holomorphic quantities, such as the superpotential or gauge kinetic terms, for
${\cal N} =2$ it extends much further as  in the famous Seiberg-Witten  solution \cite{SW}.  Finally,  for ${\cal N} =4$,  we have the much celebrated ADS/CFT correspondence \cite{ADSCFT}  between a gauge theory on the boundary and a gravity theory in the bulk of anti de-Sitter space.

In view of the above discussion, one may argue that  combining   large-$N$ ideas and supersymmetry  should provide as much an analytic predictive power as one may hope for.  However, both  large-$N$ and supersymmetry appear to take us further and further away from our initial goal:
a non-perturbative understanding of  ($N=3$, non-supersymmetric) QCD.

In the first part of this contribution we will summarize work done during the past three years on a new large-$N$ expansion that has the virtue of connecting (a sector of) ordinary QCD  to a supersymmetric
``cousin", and thus of allowing us to derive both qualitative and quantitative properties of the former.
In the second part we apply this new large-$N$ methodology to quantum mechanical gauge systems similar
to those obtained by  dimensional reduction of fully space-extended field theories.
In particular, we
define and solve (numerically as well as  analytically)
a seemingly simple supersymmetric system, which exhibits several  intriguing features such as a  phase transition
in the 't Hooft coupling and a new form of strong--weak duality.
 This should
provide, eventually, a  good and simple laboratory for testing the  idea of planar equivalence
 \cite{Armoni:2003gp}. It should also allow an extension of  recent studies of Supersymmetric Quantum Mechanics \cite{JW} to the $N \rightarrow \infty$ limit,  and to make a  possible  contact with  M-theory \cite{BFSS}.

In a sense, therefore, the two parts of this contribution to Adriano's fest try to fit with the two keywords in its title: {\it Physics} and {\it Beauty}. Achieving their synthesis represents, needless to say,  our  (yet unattained) final goal.

\section{Large-$N$ expansions: Who needs another one?}

As we mentioned in the introduction, very few techniques are
currently available for analytic studies  of
(non-supersymmetric) gauge theories such as QCD at a non-perturbative level. Among the
most promising ones, large-$N$ expansions play a special role, particularly  because of  their conjectured connection to string
theories (see e.g. \cite{Polyakov}).

The simplest and  oldest  large-$N$
expansion in QCD is the one suggested  in 1974 by 't Hooft \cite{'tH}.
It considers the limit $N_c \rightarrow \infty$ while keeping the
 't Hooft coupling
$\lambda \equiv g^2 N_c$,  as well as the number of quark flavours $N_f$,
fixed.   Only quenched planar
diagrams survive to  leading order.
Quark loops are suppressed by a power of $(N_f/N_c)$ per loop,  while non-planar diagrams are suppressed by a factor of $1/N_c^2$ per
 handle. This 't Hooft expansion led to a number of
notable successes
in  issues such as justifying the validity of the OZI   rule and in the $\eta '$ mass formula \cite{WV}.
Unfortunately, nobody  succeeded in fully solving QCD even to
leading order in
this expansion.

For  questions where  quark loops are important, e.g.  in processes with a large number of
produced hadrons,
a better approximation to full QCD is provided by the
topological expansion (TE) \cite{TE}, where  $N_f/N_c$,  rather than $N_f$ itself, is  kept
fixed in the large-$N_c$ limit. Thus, at  leading
order, the TE keeps {\em all} planar diagrams, including quark loops.
This is
easily seen by  slightly  modifying   \cite{TE}   't Hooft's
double-line notation by adding
a flavour line  to the single colour line for quarks. In the leading
(planar) diagrams the
quark loops are ``empty" inside, since gluons do not couple to
flavour.
Needless to say,  obtaining analytic results in the  TE is even harder
than in
the  case of  the original  't Hooft  expansion.

In  Ref. \cite{Armoni:2003gp}  a new
large-$N_c$ expansion, which shares some  advantages with the  TE  while
retaining  a significant predictive power, was proposed (for a review, see \cite{Armoni:2004uu}).
Its  basic idea is actually quite simple. Let us start from ordinary ($N_c=3$) QCD with $N_f$
quark flavours.
Quarks can be described by a Dirac field transforming in the
fundamental representation of SU(3)$_{\rm colour}$, or, {\em
equivalently}, in
the two-index antisymmetric representation (plus their complex
conjugates).
In extrapolating from $N_c=3$ to arbitrary $N_c$, the former alternative
leads to the 't Hooft limit. The new expansion explores instead the latter alternative,
by representing the quark of a given flavour by  a Dirac field in the  {\em
two-index antisymmetric} representation. In that case,
taking the $N_c \rightarrow \infty$  limit  at $g^2N_c$ and $N_f$ fixed
does {\it not} decouple the quark loops since, for large $N_c$, the number of
degrees of freedom in the antisymmetric field scales like the one in the adjoint,  i.e. as $N_c^2$.
For reasons explained in  \cite{Armoni:2003gp},
it  has been referred to as the orientifold large-$N_c$ limit.
The leading order of this new expansion corresponds to
the sum of all planar diagrams, in the same way as in TE,
but with the crucial difference that quark loops are now ``filled",
because the second line in the fermion propagator
is  also, now, a colour line.

The orientifold large-$N_c$ limit is, therefore, unquenched.
Its 't Hooft-notation diagrams  look, modulo  reversal of
some arrows in the fermion loops,  precisely as those of an
SU($N_c$) gauge theory with $N_f$
Majorana
fields  in the adjoint representation, a theory that we may call {\em adjoint QCD}.
Adjoint QCD can be seen, in turn, as a softly broken version of
supersymmetric Yang--Mills theory (SYM)
with $N_f -1$ additional adjoint chiral superfields.
The soft-breaking terms are just large mass terms for
the scalar fields that make them
decouple.

\subsection{Proofs of planar equivalence}
The planar equivalence between ${\cal N} =1$ SYM theory
and orientifold field theories amounts to the following statement:
an SU($N_c$) gauge theory with two Weyl
fermions in the two-index antisymmetric representation
(i.e. one Dirac antisymmetric fermion)
  is equivalent, in a bosonic subsector and as  $N_c\to\infty$ , to
  \None gluodynamics.

In Ref.~\cite{Armoni:2003gp}, a perturbative proof of the planar
equivalence was provided and a non-perturbative extension  was
outlined. In a subsequent paper \cite{Armoni:2004ub} a refined
non-perturbative proof of
 planar equivalence (extended to $N_f>1$) was given.
Its basic idea is the comparison
of   generating functionals of   appropriate   gauge-invariant  correlators in the
parent and daughter theories. This is done by,  first integrating out their respective fermions in a
fixed gauge background  and,  then, averaging over the gauge field itself.
The first step produces a fermionic determinant, which, in turn, can be expanded as a sum of Wilson loops
computed for different representations. Using some group-theory identities as well as certain factorization properties of Wilson loops at large $N$, it was possible to relate the generating functionals
of the parent and daughter theories for a subset of suitably identified sources.
Recently, a lattice version of the non-perturbative proof of \cite{Armoni:2004ub} has been
given \cite{Patella}, together with suggestions on how to check the equivalence by
realistic  lattice simulations.

It is worth noting that the two theories are not {\em fully} identical.
In particular the colour-singlet spectrum of
the orientifold theory consists  only of bosons and does not
include composite fermions at $N_c\to\infty$.

\subsection{SUSY relics in QCD}
We  will now use planar equivalence to make predictions \cite{Armoni:2003fb}
for  one-flavour QCD, keeping in mind that they are expected to be
valid up to corrections of the order of $1/N_c =1/3$ (barring
large numerical coefficients):

(i) Confinement with a mass gap. Here we assume that large-$N_c$ \None
gluodynamics is a confining theory with a mass gap. Alternatively,
if we start from the statement that one-flavour QCD confines,
we arrive at the statement that  \None SYM theory shares that property,
while the mass gaps are dynamically generated in both theories.

(ii) Degeneracy in the colour-singlet bosonic
spectrum. Even/odd parity mesons (typically mixtures of fermionic and
gluonic colour-singlet
  states) are expected to be degenerate. In particular,
\beq
{m^2 _ {\eta'} \over m^2 _ \sigma}=1 + O(1/N_c)
\,, \qquad\mbox{one-flavour QCD}\,,
\label{etasigma}
\eeq
where $\eta '$ and $\sigma$ stand for $0^-$ and $0^+$
mesons, respectively.
This follows from the exact degeneracy in \None SYM theory. Note that the
$\sigma$ meson is stable in this theory, as there are no pions.
This prediction  should be taken with care (i.e.
a rather large numerical coefficient in front of $1/N_c$ may occur), since
the
$\eta'$ mass is given by the anomaly (the WV
formula \cite{WV}), whereas the $\sigma$
mass is more ``dynamical.'' The
degeneracy between even and odd-parity mesons should improve at higher
levels on the expected Regge trajectory.
In order to check this relation in {\em real QCD}, let us assume
that the $\sigma$ mass is not very sensitive to the number of flavours.
On the other hand, according to the WV formula
\cite{WV} and neglecting quark masses, the
$\eta'$ mass scales like $\sqrt{N_f}$; we can therefore
 extrapolate relation \eqref{etasigma} to
obtain a prediction for real QCD
\beq
m_{\eta '} \sim \sqrt 3 m_{\sigma}.
\eeq
  Although the  $\sigma$ bump is very broad, it is encouraging that the above
relation is indeed in qualitative  agreement with the position of the
enhancement in the appropriate $\pi\pi$ channel.

(iii) Bifermion condensate. \None SU$(N_c)$ gluodynamics has a
bifermion condensate  \cite{gluinocond} that can take $N_c$ distinct
values:
\beq
\langle \lambda  \lambda \rangle_k  \sim M_{\rm uv}^3\,  e^{- \tau
/N_c } e^{2 i \pi k/N_c} =
c \Lambda^3 \, e^{i(\theta +2 \pi k)  /N_c} \, , \,\,\,\; k = 0, 1,
\dots, N_c-1 ,
\label{bfcsusy}
\eeq
with
$$\tau =\frac{8\pi^2}{g^2} - i\theta\,$$
and $c$ a calculable numerical coefficient.
  The finite-$N_c$ orientifold field theory is  non-supersymmetric,
and here we expect (taking account of pre-asymptotic $1/N_c$
corrections) $N_c-2$ degenerate vacua with
\begin{eqnarray}
\langle \bar \Psi _L \Psi _R\rangle_{k'}  &\sim& M_{\rm uv}^3\,
\exp\left\{-
{8\pi ^2\over
g^2 (N_c+4/9)} + i{ \theta + 2 \pi k'\over {N_c-2}}\right\}
\nonumber\\[3mm]
&\sim&
c' \Lambda^3 \, \exp\left\{i{\theta + 2 \pi k'\over N_c-2}\right\}
\, , \quad k' = 0,1, \dots , N_c-3\,.
\label{bfco}
\end{eqnarray}
The term $4/9$ in \eqref{bfco} is due to the one-loop
$\beta$ function of the orientifold
field theory,
$b=3N_c +{4\over 3}$, while $N_c-2$ in \eqref{bfco} is twice the dual
Coxeter number of the
antisymmetric representation (fixing the coefficient of the axial
anomaly).
Finally, $c'$ is a normalization factor that we
will discuss below.

\subsection{Analytic estimate of the quark condensate in QCD}

The relation between the quark condensate in
one-flavour QCD and the gluino condensate in SYM theory
can be pushed further \cite {Armoni:2003yv}  by appealing to the fact that the quark condensate of
QCD$_{OR}$ must agree with the gluino condensate at $N = \infty$ and must vanish at $N=2$.
Thus, its value at $N=3$ (which is nothing but the quark condensate of one-flavour QCD) can be obtained by {\it interpolating} between these two values.
The final outcome of such an analysis \cite {Armoni:2003yv} is the following analytic formula:

\beq
\langle \bar \Psi \Psi \rangle_\mu = - {3 \over 2\pi^2} \mu ^ 3
(\lambda(\mu))^{-{\gamma \over \beta _0} - {3\beta _1 \over \beta _0 ^2}}
\exp\biggl(- {9\over \beta_0 \lambda (\mu)}\biggr) k(1/3) \, ,
\label{condensate}
\eeq
where   $\mu$ is the renormalization scale for the condensate,
$\beta_0$ and $\beta_1$ are the one- and two-loop $\beta$-function coefficients, $\gamma$ is the one-loop anomalous dimension of the condensate, and $k(1/N)$ stems for further,  unidentified $1/N$ corrections.

In a recent paper \cite{Shore:2005} an attempt was made to extend these considerations to an arbitrary
number of quark flavours. The idea (actually first proposed for different reasons in \cite{Corrigan:1979xf}) is to add to QCD$_{OR}$ $n_f$ quarks in the (traditional) fundamental representation of SU($N$), so that they become irrelevant in the large-$N$ limit.
This new class of theories has been dubbed QCD$_{OR'}$.
At $N=3$, such a theory is nothing but QCD with $N_f$ flavours where $N_f = n_f +1$.
One has to determine  which sectors of QCD$_{OR'}$ can be mapped into corresponding quantities of SYM theory, in particular since the former theory has a NG sector which is absent in SYM.
When this is done \cite{Shore:2005} one obtains an interesting generalization of (\ref{condensate}) to $N_f > 1$ and can even argue that the connections should be particularly accurate for the ``realistic" case of $N_f = 3$.
 For three-flavour QCD the relevant values are
$\beta_0= 9,\, \beta_1=32,\, \gamma =4$.

In order to calculate the condensate \eqref{condensate} we need to
know the value of the 't Hooft coupling $\lambda$ at a scale $\mu$
that we choose to be $\mu = 2\,{\rm GeV}$.
The Particle Data Group \cite{PDG} quotes
$\alpha_s(2 {\rm~GeV}) = 0.31 \pm 0.01$,
which corresponds to $\lambda (2~{\rm GeV}) = 0.148 \pm 0.010$.
We therefore choose to plot the function
\beq
 \frac{\langle \bar \Psi \Psi \rangle_ {\rm 2\, GeV}}{{\rm GeV}^3} =
- {3 \over 2\pi^2} 2 ^ 3
\lambda ^{- {44
\over 27} } \exp \biggl(- { 1\over \lambda }\biggr) ,
\label{value}
\eeq
in a range of $\lambda$. Comparison between the above theoretical prediction and present determinations of the condensate
is shown in Fig. 1. Clearly the result supports the validity of planar equivalence within the expected
precision at $N=3$.

\begin{figure}[h]
\begin{center}
\includegraphics[width=0.8\textwidth,clip=true,trim=20 10 175 475]{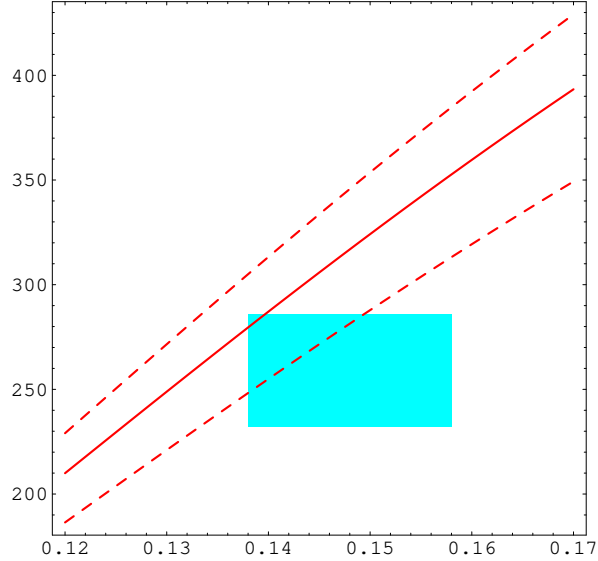}
\end{center}
\caption{The quark condensate expressed as $-(y~{\rm MeV})^3$ as a function of
the 't Hooft coupling $\lambda$. The $\pm 1\sigma$ range of the coupling,
$0.138 < \lambda < 0.158$, and the lattice estimate
$-(259 \pm 27~{\rm MeV})^3$ define the shaded region.}
\label{fig}
\end{figure}

\section{Planar Quantum Mechanics}
In this second part of the paper, we turn from QFT to the simpler case of
Quantum Mechanics and try to define a ``Planar Quantum Mechanics"  that hopefully mimicks the large-$N$ limit of a system whose degrees of freedom
are $N \times
N$ (bosonic or fermionic) matrices. The hope is that, eventually, this technique can be extended to QFT and  provide a new handle on  solving interesting gauge theories --such as YM, SYM and QCD--  in the large-$N$ limit.
It may also provide useful information on whether (and how) ``planar equivalence" may or may not  work.

\subsection{Fock states, and matrix elements at large $N$}
Our quantum mechanical system is defined by a Hamiltonian $H$, which is  polynomial in $N \times
N$-matrix  annihilation and in creation
operators $a_{ij}$, $\ad_{ij}$. Symmetry under the $U(N)$ transformations is ensured by taking $H$ to be the trace of such a polynomial. Creation and annihilation operators satisfy standard commutation relations
\eqn
[a_{ij},\ad_{kl}]=\delta_{il}\delta_{jk}  \label{com} \, .
\eqnx
We shall work in the eigenbasis of the ${\rm U}(N)$-invariant number operator $B={\rm Tr}(\ad a)$.
Therefore basis states are constructed by acting on the Fock vacuum with the invariant building blocks,
or ``bricks", ${\rm Tr}[(\ad)^n]$, $n=1,2, \dots, N$. Following the Cayley--Hamilton theorem  there are $N$ independent
bricks for the ${\rm U}(N)$ group.

A general state is a product of all possible bricks and their powers. It will prove useful to define a cut
Hilbert space $H_B$,
i.e. a space restricted to states with no more than $B$ bosonic quanta. Such a space is spanned by all polynomials
of bricks with maximum $B$-th order. Obviously there are many states with a fixed number of bosonic quanta $n$.

In the large-$N$ limit,  things simplify \cite{VW1} considerably: for given $n$,  there is only one relevant state.
It is created by  the single-trace operator:
\eqn
|n \rangle =\frac{1}{{\cal N}_n} Tr[(\ad)^n]|0\rangle \, ,  \label{state}
\eqnx
where ${\cal N}_n$ is a suitable normalization factor.
All states that are created by products of traces are non-leading in the sense  that they give rise
to non-leading matrix elements. A second simplification is that leading operators
also have a single trace form.
 These two basic rules of the Planar Calculus are best illustrated by the explicit examples below.

\noindent {\em Normalization of planar states}
\newline
 The normalization factor  ${\cal N}_n$ in (\ref{state}) reads
\eq
{\cal N}_n^2= \langle 0|Tr[a^n] Tr[(\ad)^n]|0 \rangle = \langle  0|(12)(23)...(n1)[1'2'][2'3']\dots[n'1']|0 \rangle,  \label{norm}
\eqx
where we have simplified the notation, e.g. $a_{i_1,i_2}\rightarrow (12)$ and $\ad_{i_7,i_8}\rightarrow [7'8']$
and all indices are summed. Using (\ref{com}) it is easy to see that the maximal power of $N$ is achieved
by contracting indices between () and [] starting from the middle and working outwards. This produces the famous
planar diagrams. One way to picture it  is to represent ()()...() as a circle with $n$ pairs of short lines,
representing indices,
pointing inside and [][]...[] as a smaller circle with lines pointing outside.  Leading contributions result
when opposite pairs of lines match: this
gives $n$ loops, i.e. a factor $N^n$. Because of the cyclic properties of a trace in (\ref{norm}), one gets $n$ such
contributions.
Moreover, according to (\ref{com}), each contraction gives a factor $1$. The final result reads
\eq
{\cal N}_n=\left\{
\begin{array}{cc}
\sqrt{n} N^{(n/2)}, & n > 0, \\
1, & n = 0.
\end{array}\right.
\eqx

\noindent {\em Example of matrix elements}
\newline
 As another example, we calculate the matrix element of a typical term in a generic Hamiltonian:
 \eq
 H_{n+2,n}=g^2  \langle n+2|Tr[\ad\ad\ad a]|n\rangle.  \label{hmn}
 \eqx
Consider a state
\eqn
Tr[\ad\ad\ad a]Tr[(\ad)^n]|0\rangle&=&[12][23][34](41)[1'2'][2'3']...[n'1']|0\rangle \nn \\ &=&
n [1'2][23][32'][2'3'][3'4']...[n'1']|0\rangle,
\eqnx
$n$ equal terms result from commuting one annihilator, (41), all the way until it hits the vacuum state.
Each  contraction (\ref{com}) gives just a factor $1$.
Pictorially, one can attach a small circle with 3 creators and one
annihilator to a big circle with $n$ pairs of indices in $n$ equivalent ways. The resulting state is
proportional to $|n+2\rangle$. Collecting the normalization factors gives
\eqn
H_{n+2,n}&=&  g^2 n \frac{1}{{\cal N}_n}  \langle n+2|Tr[(\ad)^{(n+2)}]|0\rangle
         = g^2 n \frac{\sqrt{(n+2)/n}N}{ {\cal N}_{n+2}} \times \nn \\
         &&  \langle n+2|Tr[(\ad)^{(n+2)}]|0 \rangle
 = g^2 N \sqrt{n(n+2)} ,\;\;\; n > 0.
\eqnx
Indeed, such an interaction term depends only on $\lambda = g^2 N$, showing the  relevance of the t'Hooft  coupling.
To complete this example, we also calculate a matrix element of the conjugate operator
 \eq
 H_{n-2,n}=g^2  \langle n-2|Tr[\ad a a a]|n\rangle=\frac{1}{{\cal N}_n}
 \langle n-2|{\rm Tr}[\ad a a a]{\rm Tr}[(\ad)^n]|0\rangle.  \label{chmn}
 \eqx
The leading contribution comes from the three subsequent contractions of adjacent $a's$ and $\ad's$, which give
a factor $N^2$. Again this can be done in $n$ ways.
Taking into account the renormalization ${\cal N}_n\rightarrow {\cal N}_{(n-2)}$
gives finally
\eqn
H_{n-2,n}&=&g^2 N \sqrt{n(n-2)},\;\;\; n > 2,
\eqnx
which again shows the  t'Hooft scaling and, upon the replacement $n\rightarrow n+2$, confirms
the hermiticity of the planar hamiltonian.

\subsection{An anharmonic oscillator}
Let us now apply planar rules to find the spectrum of the $U(\infty)$-invariant anharmonic oscillator.
The Hamiltonian
\eqn
H= {\rm Tr}[p^2]+g^2  {\rm Tr}[x^4] \equiv T + V  \label{H1}
\eqnx
is bounded from below and represents a perfectly well defined system {\em per se}. On the other hand,
there exists a whole family of similar Hamiltonians (supersymmetric or not), which have been obtained
from gauge-invariant field theories by reducing the system to a single point in space.
For example, reduced  two-dimensional Yang--Mills gluodynamics is described solely by the kinetic
term of Eq. (\ref{H1}) with a global $U(N)$ invariance as the reminder of the local
gauge symmetry of the original, space-extended  theory.

As a first step, introduce matrix creation and annihilation
operators
\eqn x=\frac{1}{\sqrt{2}}(a+\ad),&& p=\frac{1}{\sqrt{2}
i}(a-\ad),
\eqnx
and rewrite each term of $H$ in the normal-ordered
form:
\eqn
H&=&\frac{1}{2}N^2+\frac{1}{2}N^3 g^2 \nn \\
&+& \left(-\frac{1}{2}+N g^2\right)(\ad^2+ a^2)+
(1+2 g^2 N)\ad a \nn \\
&+& \frac{g^2}{4}(\ad^4+a^4)+g^2 \ad^2 a^2 + g^2(\ad^3 a + \ad
a^3) \label{ahex}
\eqnx
A note on the normal ordering may be useful here. For higher powers of a's and $\ad$'s the normal ordering may
spoil the group structure. For example
\eq
:a_{ik}\ad_{kj} a_{jl} \ad_{li}: = \ad_{kj}\ad_{li} a_{ik} a_{jl}, \nn 
\eqx
is not a trace, hence it does not contribute to the planar limit.
On the other hand
\eq
:\ad_{ik} a_{kj} a_{jl} \ad_{li}: = \ad_{ik}\ad_{li} a_{kj} a_{jl}, \nn 
\eqx
can be brought to the trace form - it preserves the group structure - and consequently
contributes to the leading behavior. As an example consider one of the quartic terms
in the expansion of the potential in Eq.(\ref{H1}).
\eqn
(2,2)&=& Tr[a^2 \ad^2 + a\ad a\ad+a\ad\ad a +\ad a a \ad + \ad a \ad a + \ad^2 a^2]. \nonumber
\eqnx
Commuting all annihilation operators to the right and retaining only single traces gives
\eqn
(2,2)&=& 2 N^3+ 8 N Tr[\ad a] + 4 Tr[\ad^2 a^2].
\eqnx
Out of the six quartic terms only four preserve the trace structure after bringing them to the
normal ordered form. The remaining two do not, hence they are non-leading.

The calculation of all matrix elements of (\ref{ahex}) closely follows  the earlier examples.
After some algebra one obtains
\eqn
<n|H|n>&=&\frac{N^2}{2}(1+\lambda)+ \left\{
\begin{array}{cc}
(1+3\lambda)n, & n \geq 2, \\
(1+2\lambda)n, & n < 2,
\end{array}\right.\nonumber \\
<n+2|H|n>&=<n|H|n+2>&=\left\{
\begin{array}{cc}
\left(-\frac{1}{2}+2\lambda\right)\sqrt{n(n+2)}, & n > 0, \\
\left(-\frac{1}{2}+\lambda\right)\sqrt{2}N & n = 0,
\end{array}\right.\label{Hanoex}\\
<n+4|H|n>&=<n|H|n+4>&=\left\{
\begin{array}{cc}
\frac{1}{4}\lambda\sqrt{n(n+4)}, & n > 0, \\
\frac{1}{2} \lambda N & n = 0.
\end{array}\right. \nonumber
\eqnx
This result illustrates an important feature of all large-$N$ calculations (also valid in QFT).
Namely, not all matrix elements scale with $\lambda$. There exists a class of "superleading"
contributions which are divergent in the 't Hooft limit. They result from  vacuum
diagrams and should be treated separately \footnote{We are grateful to E. Onofri for useful
discussions on this issue.}. At the moment, we just neglect them, i.e. we use
in the following
the subtracted Hamiltonian $H-N^2(1+\lambda)/2$ and ignore the first row and column
where the non-scaling vacuum diagrams contribute only.

We may now go ahead and diagonalize the so modified Hamiltonian $\tilde{H}$. To
this end we introduce a cut-off $ n \le B $, find numerically the
spectrum of $\tilde{H}$, and increase $B$ until the convergence is
reached \cite{CW}.

\begin{figure}[p]
\begin{center}
\includegraphics[width=1.0\textwidth,clip=true,trim=40 20 100 200]{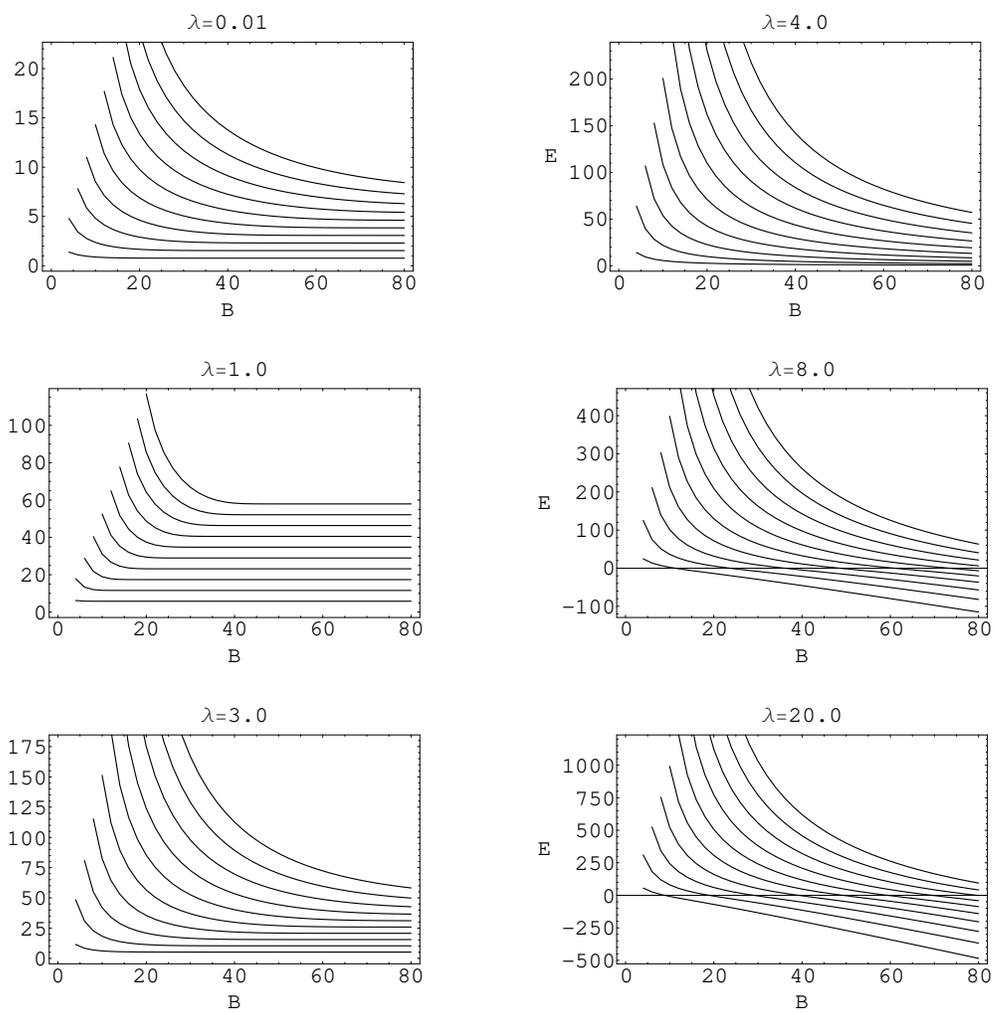}
\end{center}
\caption{Cut-off dependence of the spectrum of the anharmonic oscillator in the even-parity sector.}
\label{allNO}
\end{figure}

Our results are shown in Fig. \ref{allNO} and can be summarized as
follows.

\begin{itemize}
 \item Satisfactory convergence of lower eigenenergies,
with increasing cut-off,  is achieved for $\lambda < \;\; \sim4$. In this region
the system resembles an effective harmonic oscillator with almost (but not exactly) equidistant
 "infinite volume" levels. This seems to be a general feature of the large-$N$ dynamics \cite{MO}~
 \cite{MP}.
 \item With $\lambda$ approaching $\lambda_c\sim 4$ the mass gap vanishes and all levels
 fall towards zero with increasing cut-off. Such a behaviour has been studied before and
 is the characteristic sign of the continuous spectrum in the infinite-volume limit \cite{TW}.
\item For $\lambda > \lambda_c$ all levels fall with the cut-off towards $-\infty$; $\bar{H}$
  appears to be unbounded from below.
\end{itemize}
The last point may seem unnatural, signalling  some trouble for the planar approach. In fact
it provides one more argument that a quantum mechanics turns into a field theory in the large-$N$ limit.
Namely, bound states with negative energies are of course quite common in both systems. However
only a field theory, with its multiparticle sectors, can account for an infinite series of negative states.\footnote{JW thanks P. Menotti for a very instructive discussion on this subject.}

Nevertheless this emergence of the negative eigenenergies provides an important lesson (and a warning)
about possible intricacies
of the planar limit. The original Hamiltonian (\ref{H1}) is perfectly well defined and positive for any
finite $N$. This raises the question of the relation between the "true" large-$N$  limit, defined
as solving non-perturbatively a system at finite $N$ and {\em then} taking $N$ large, and the planar limit
discussed here. There may be many reasons for the non-commutativity of the two, the most obvious
one being the {\em ad hoc} subtraction of superleading contributions.
The model discussed hereafter does not have this deficiency.

\subsection{A very symmetric supersymmetric system}

We now add fermionic degrees of freedom described by matrix creation and annihilation operators
$\fd,\;f,$ $\{f_{ij} \fd_{kl}\}=\delta_{il}\delta_{jk}$ .
To avoid vacuum diagrams, we require that supersymmtery charges (hence also a Hamiltonian) annihilate
the empty Fock state. A simple choice is \cite{VW1}
\eqn
Q= {\rm Tr} [f \ad(1+g\ad)],
\;\;\; \Qd= {\rm Tr} [\fd (1+g a) a]\;\;\;H=\{Q,\Qd\}=H_B+H_F,
\eqnx
or explicitly (a trace is always implied)
\eqn
H_B&=&\ad a + g(\ad^2 a + \ad a^2) + g^2 \ad^2 a^2 , \nn \\
H_F&=& \fd f + g ( \fd f (\ad+a) + \fd (\ad+a) f) \label{susyH} \\
& + & g^2 ( \fd a f \ad + \fd a \ad f + \fd f \ad a + \fd \ad f a) . \nn
\eqnx
This Hamiltonian conserves the fermion number $F=\fd f$. In each fermionic
sector the planar basis $\{|F,n \rangle \}$ is now created by the single trace with $F$ fermionic and $n$ bosonic
creation operators. It is now a simple, but somewhat tedious, exercise to calculate the matrix elements
of $H$ in this representation. We obtain for the first two fermionic sectors:
\eqn
\langle 0,n|H|0,n \rangle&=&(1+\lambda)n -\lambda \delta_{n1}, \nn \\
\langle 0,n+1|H|0,n \rangle=\langle 0,n|H|0,n+1 \rangle&=&\sqrt{\lambda}\sqrt{n(n+1)}, \label{susyHF0}\\
\langle 1,n|H|1,n \rangle&=&(1 + \lambda)(n+1) + \lambda, \nn \\
\langle 1,n+1|H|1,n \rangle=\langle 1,n|H|1,n+1 \rangle&=&\sqrt{\lambda}(2+n). \label{susyHF1}
\eqnx

\begin{figure}[tbp]
\includegraphics[width=1.0\textwidth,clip=true,trim=10 20 200 300]{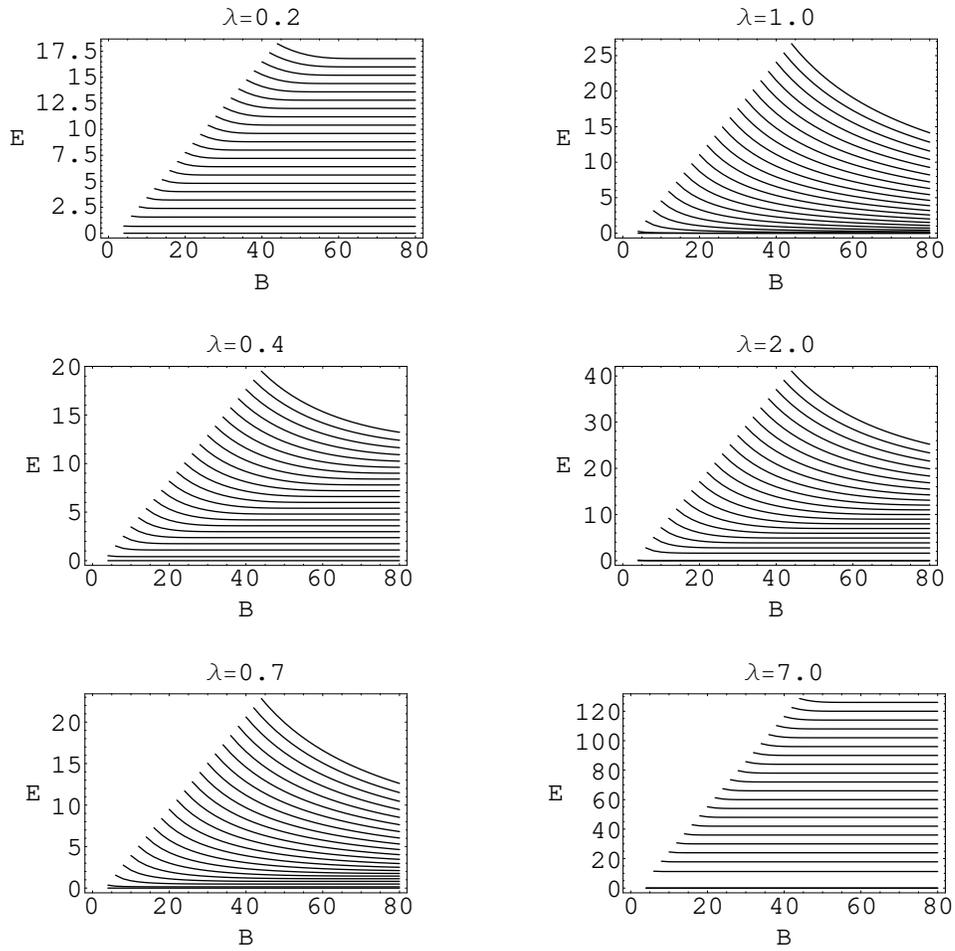}
\vskip-4mm
\caption{Cut-off dependence of the spectrum of  $H$, in the $F=0$ sector,
in a range of values for $\lambda$.}
\label{fig:h2}
\end{figure}

The numerical calculation of the spectrum proceeds in the same way as for the anharmonic oscillator
(see Fig.~\ref{fig:h2}).

\begin{figure}[tbp]
\begin{center}
\epsfig{width=1.0\textwidth,trim=-100 00 200 600,file=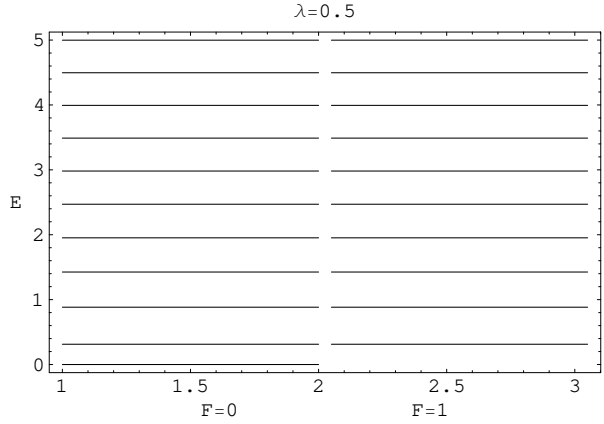}
\end{center}
\vskip-4mm \caption{First 10 energy levels of $H$, in $F=0$ and $F=1$
sectors, at $\lambda=0.5$. } \label{fig:susyf01}
\end{figure}

\begin{figure}[tbp]
\begin{center}
\includegraphics[width=0.8\textwidth,clip=true,trim=40 00 200 250]{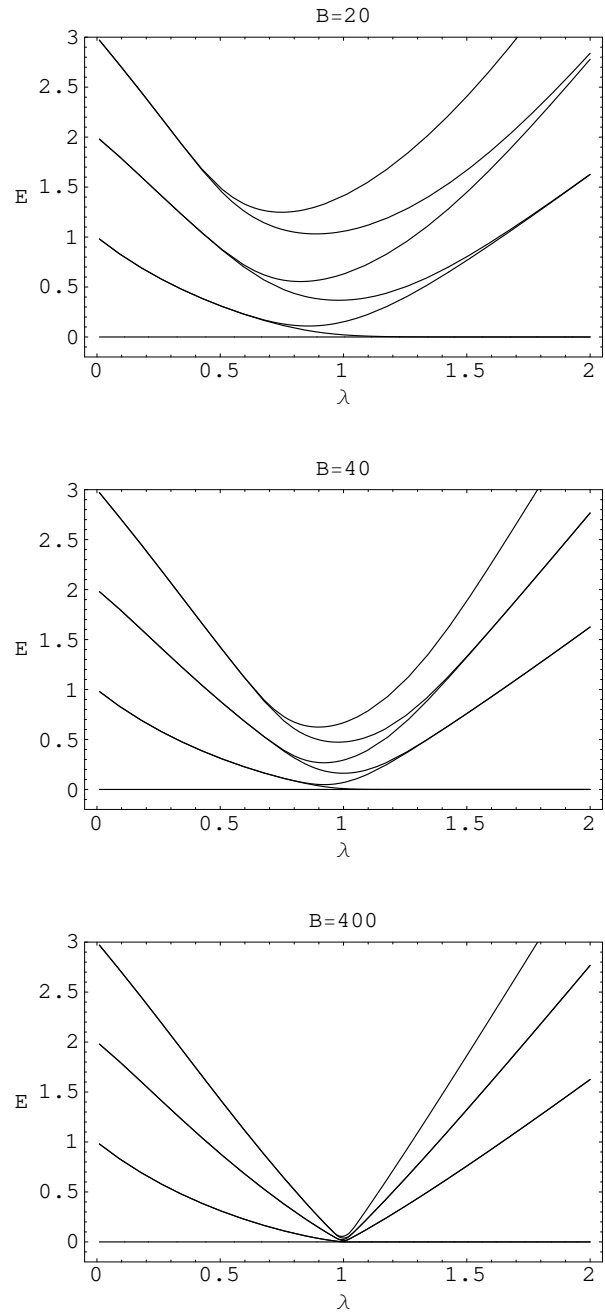}
\end{center}
\vskip-4mm
\caption{Rearrangement of the $F=0$ and $F=1$ spectra, around $\lambda=1$,
for increasing cut-offs.}
\label{fig:rearrange}
\end{figure}

Now, however, the salient features of this system are:
\begin{itemize}
\item The spectrum is non-negative for all values of the 't Hooft coupling. This confirms the suspicion
of vacuum
diagrams being responsible for the earlier problems.
\item For $\lambda$ away from 1 the eigenenergies (at infinite cut-off) are again almost
equally separated with a common mass gap depending on $\lambda$.
\item There is a phase transition at $\lambda_c=1$. At this point, and only there, the spectrum loses its mass gap and becomes continuous.
\item In the vicinity of  $\lambda_c$ we observe a critical slowing down: higher cut-offs are required
to achieve satisfactory convergence.
\item  There is an excellent boson--fermion degeneracy, see  Fig.~\ref{fig:susyf01}.  Supersymmetry is unbroken at infinite cut-off but is quite badly broken at finite cut-off near the critical point $\lambda_c$.
\item In both phases there is an unpaired bosonic SUSY vacuum with zero energy. This is the empty Fock state, which, by construction, is annihilated by our Hamiltonian.
\item While moving across the transition point, at finite cut-off, members of supermultiplets rearrange; see Fig.~\ref{fig:rearrange}. In particular
{\em a second} ground state with zero energy appears in the strong coupling phase.
\item Some symmetry between the strong and weak coupling regions emerges.
\end{itemize}

It turns out that the problem can be solved analytically, offering better understanding of these features.
We shall now discuss each one of  them in turn.

\subsubsection{Supersymmetry}

With the rules of Sect.2  the matrix representation of supersymmetry charges is easily obtained in the planar limit:
\eqn
\langle 1,n-1|\Qd|0,n\rangle = \sqrt{n},\;\;\;\langle 1,n-2|\Qd|0,n\rangle = 0,n\rangle = b \sqrt{n}.
\label{QdM}
\eqnx
Obviously the $\Qd$ generator has matrix elements only between $F=0$ and $F=1$ sectors. One can now readily
check the sum rule
\eqn
\sum_m \langle 0, n' |Q | 1,m\rangle \langle 1,m |\Qd | 0,n\rangle  = \langle 0,n' |H | 0,n\rangle  \label{sr1}
\, ,
\eqnx
which follows from $H=\{Q,\Qd\}$. In our representation its test becomes a simple exercise
in matrix multiplication.
Notice that even the exceptional diagonal contribution at $n=1$ is correctly reproduced by the anticommutator
of SUSY charges.

A little more involved test employs the identity
\eqn
\langle 0,n' |H^2 | 0,n\rangle = \langle 0,n' |Q H \Qd| 0,n\rangle,  \label{sr2}
\eqnx
which again turns into  straightforward matrix algebra. Notice that the
sum rule resulting from (\ref{sr2}) relates matrix elements of $H$ in {\em different} fermionic sectors.

 We conclude that the planar approximation preserves supersymmetry. On the other hand the cut-off in terms
 of the number of bosonic quanta  explicitly breaks SUSY. Consequently the fermion--boson degeneracy is observed
 numerically only at sufficiently large cut-offs. It is conceivable that  SUSY invariant cut-offs can be found
 \cite{CW1}.

\subsubsection{The second ground state}
\begin{figure}[tbp]
\begin{center}
\epsfig{width=1.0\textwidth,trim=50 000 100 650,file=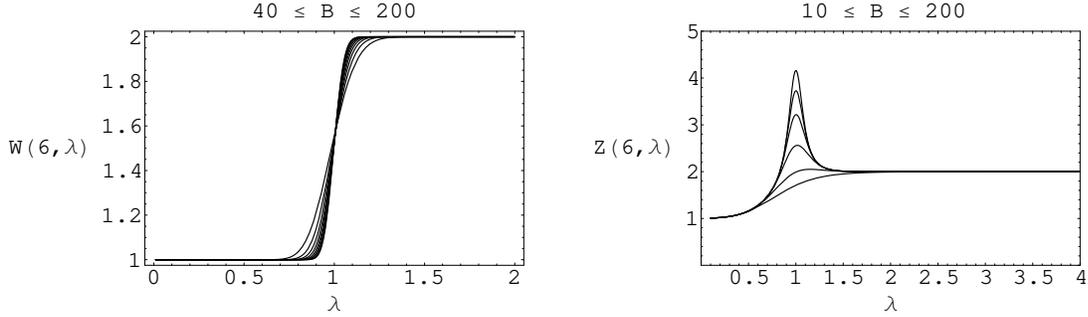}
\end{center}
\vskip-4mm \caption{$\lambda$ dependence of the Witten index and
the partition function, at $\beta = 6$, for different cut-offs.}
\label{wz}
\end{figure}

To understand the existence of the second massless state,
let us introduce the composite creation and annihilation operators $\ad_n$ and $a_n$, which
create the planar basis in the new Hilbert space:
\eqn
|0,n\rangle=\ad_n |0\rangle.
\eqnx
In terms of these operators the $F=0$ Hamiltonian reads
\eq
 H^{(F=0)} = \ad_1 a_1 + \sum_{n=1}^{\infty} n(1+ b^2) \ad_n a_n +
\left( \sum_{n=1}^{\infty} b \sqrt{n(n+1)} \ad_n a_{n+1} + {\rm h.c.} \right) \ .
\eqx
Then one can formally construct the following state
\eqn
|0\rangle_2 = \sum_{n=1}^{\infty} \left(\frac{-1}{b}\right)^{n} \frac{\ad_n}{\sqrt{n} }|0\rangle_1 \, ,
\label{vac2}
\eqnx
which is annihilated by $Q^{\dagger}$, Eq. (\ref{QdM}), hence also by the Hamiltonian (\ref{susyHF0}).
It is normalizable only for $\lambda >1$, thereby explaining the puzzle found numerically.
The emergence of this second ground state in the strong coupling phase causes the Witten index \cite{WQM}
to jump by one unit (in the sectors with $F=0,1$) across the phase transition. According to the
Feynman--Kac relation,
the thermal partition function also reveals such a discontinuity at zero temperature. However, since
 higher levels collapse to zero at $\lambda_c$, there is also a $\delta$-like contribution
at this point:  c.f.  Fig.~\ref{wz}.

\subsubsection{Strong--weak duality}
The qualitative symmetry of the energy levels between the strong and weak coupling phases,
seen in Fig.~\ref{fig:rearrange},
finds its explanation in a rather intriguing exact duality between the two regimes. Consider the Hamiltonian
(\ref{susyHF1}) in the $F=1$ sector. It is evident that the ``reduced" Hamiltonian
\eqn
\bar{H}= \frac{1}{b}(H-b^2)   \label{dualF0}
\eqnx
is symmetric under the replacement $b\rightarrow1/b$. Therefore the eigenenergies of $H$ satisfy
\eqn
 \frac{1}{b} \left( E^{(F=1)}_n(b)   -b^2 \right) = b \left(E^{(F=1)}_n(1/b) -\frac{1}{b^2}\right).    \nn
\eqnx
The same reciprocity relation in the $F=0$ sector {\em does not} hold at  the operator level, c.f. (\ref{susyHF0}).
On the other hand, because of the supersymmetry, the duality works there as well.
However, since there exists a second bosonic vacuum, in this sector duality
relates eigenenergies with different indices:
\eqn
\frac{1}{b} \left( E^{(F=0)}_{n}(b) -b^2 \right) = b \left(E^{(F=0)}_{n+1}(1/b) -\frac{1}{b^2}\right),\;\;\; b<1.
 \label{DR}
\eqnx
\subsubsection{The analytic  solution}
Finally, and rather surprisingly, it turns out that the Hamiltonian (\ref{susyHF0})
can be analytically diagonalized. More details can be found in \cite{VW1}; here we
outline only the main steps of the derivation.

1.  The $F=0$ Hamiltonian is strikingly simple when written
in terms of the linear combinations of $a_n$:
\eqn
H = \sum_{n=1}^{\infty} B^{\dagger}_n B_n\; , \;\;\;
B_n&=&\sqrt{n} a_n + b \sqrt{n+1} a_{n+1} \, , \nonumber \label{htwo}
\eqnx

2. The states created by the $B_n^{\dagger}$ operators
\eqn
  |B_n\rangle \equiv B_n^{\dagger}|0\rangle&=&\sqrt{n}|n\rangle+b\sqrt{n+1}|n+1\rangle \, .
\eqnx
form a non-orthonormal basis.
 We shall find the eigenstates $|\psi \rangle $ of the reduced Hamiltonian by expanding them into the
$|B_n\rangle$ basis
and constructing the generating function $f(x)$ for the expansion coefficients:
\eqn
|\psi\rangle = \sum_{n=0}^{\infty} c_n |B_n\rangle \;\;\;\;\leftrightarrow \;\;\;\;f(x)=\sum_{n=0}^{\infty} c_n x^n . \, \;\;\;\;\;\;\
 \label{spsi}
\eqnx

3. The simple way in which $\bar{H}$ acts on the $|B_n\rangle$ basis implies the following
first-order differential equation for $f(x)$
\eq
\bar{H}|\psi\rangle=\epsilon|\psi\rangle\;\;\;\;\leftrightarrow\;\;\;\;\;
w(x)f'(x) + x f(x) - b f(0) - f'(0) = \epsilon f(x),  \label{hx}
\eqx
where  $w(x)=(x+b)(x+1/b)$, and the eigenvalues of $H$ are simply related to those of $\bar{H}$:
$E = b(\epsilon+b)$.

4. A solution of (\ref{hx}) is straightforward (see again \cite{VW1} for details). The final expression for
 the generating function reads

\eq
f(x)=  \left\{ \begin{array}{cc}
          \frac{1}{\alpha}  \;\frac{1}{x+1/b}\; F(1,\alpha;  1+\alpha;\frac{x+b}{x+1/b}),& b < 1 \\
          \frac{1}{1-\alpha}\;\frac{1}{x+b}  \; F(1,1-\alpha;2-\alpha;\frac{x+1/b}{x+b}),& b > 1
          \end{array} \right. \ , \;\;\;\;\;\alpha=\frac{\epsilon+b}{b-1/b}\; ,
\label{sol}
\eqx
where $F(a,b,c;x)$ is the standard geometric function, and the quantization condition,
\eqn
(\epsilon + b) f(0)=0 \, ,
 \label{qcon}
\eqnx
determines the discrete series of eigenenergies in the $F=0$ sector.

\begin{figure}[tbp]
\begin{center}
\includegraphics[width=0.8\textwidth,clip=true,trim=00 00 300 600]{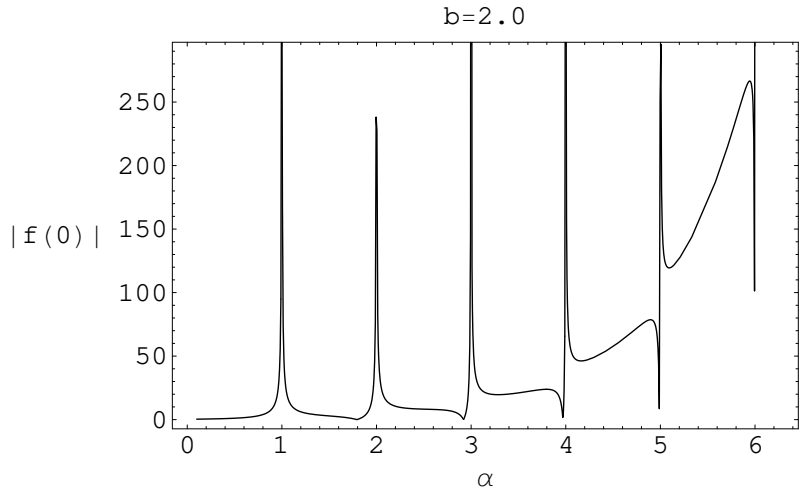}
\end{center}
\vskip-4mm
\caption{The quantization condition as a function of $\alpha$.
The first four zeros are clearly visible. To see higher zeros the $\alpha$ resolution
of the plot must be increased. }
\label{fig:piki}
\end{figure}

This analytic solution nicely explains all results discussed above. In particular:
\begin{itemize}
\item The absolute value of $f(0)$ reveals a series of zeros in $\alpha$, c.f. Fig.~\ref{fig:piki},
which agree with the large cut-off limit of the numerical eivenvalues of (\ref{susyHF0}).
\item The duality among massive eigenvalues follows immediately from (\ref{qcon}) and the symmetry of (\ref{sol}) under
the substitution $b\rightarrow1/b$ and $\alpha \rightarrow 1-\alpha$.
\item The collapse  of the eigenenergies to zero at the critical point is evident from $E = \alpha(b^2-1),$
and so is the fact that the roots $\alpha_n$ are bounded  by  nearby poles of the $\beta$ function.
\item The generating function of the second ground state can be easily obtained by setting $\alpha=0$
in Eq. (\ref{sol}) for $b>1$. This gives
\eqn
f_0(x)=\frac{1}{1+b x} \log\frac{b+x}{b-1/b}, \;\;\;  b>1,
\eqnx
which indeed corresponds to the expansion (\ref{vac2}). On the other hand, one cannot set
$\alpha=0$ in the $b<1$ solution  -- the second vacuum {\em does not} exist for $b<1$.
\end{itemize}

Given (\ref{sol}) and (\ref{qcon}) one can study the flow of the eigenvalues in both phases.
Perhaps the most interesting is the behaviour in the vicinity of the critical point. From the known
asymptotics of the hypergeometric functions around $\lambda=1$ we conclude that the first eigenvalue
tends to zero as $-(1-\lambda)/\log(1-\lambda)$ when $\lambda \rightarrow 1^-$, which results in a vanishing
first --and infinite second--  derivative.

One can also quantitatively study the free energy near the phase transition, which appears to be
stronger than in the Gross--Witten model \cite{GW}.

\section{Higher-$F$ sectors}

The structure of higher-$F$ sectors, rather than being a simple repetition of the $F=0,1$ sectors, appears to exhibit novel interesting  features.
This already follows quite directly from looking at the free theory ($\lambda =0$).
Even in that case the structure of the multiplets, and the way they should rearrange into supermultiplets,
is highly non-trivial. This is best illustrated in a plot that resembles the old  Chew--Frautschi plot  (CFP),  in which
angular momentum is plotted against squared mass\footnote{ Linearity of  the (Regge) trajectories in this plot was perhaps the first  evidence for a string-like structure of  hadrons,  with  the inverse of the Regge slope reinterpreted as the string tension.}.

As one can see, while at all mass levels there is just one $F=0$ and one $F=1$ state that naturally form
SUSY doublets, this is not the case for the $F=2, F=3$ states. Their number, at high enough occupation numbers,
is different, the excess  always being in favour of the $F=3$ sector.
However, supersymmetry must be satisfied, and this is achieved non-trivially by having some of the
$F=3$ states pair with those with $F=4$. Once more, this leaves some of the $F=4$ states unpaired but,
continuing the exercise upwards in the CFP,  shows that, eventually, every state finds its  supersymmetric partner.

\begin{figure}[tbp]
\begin{center}
\epsfig{width=1.0\textwidth,trim=-100 00 00 500,file=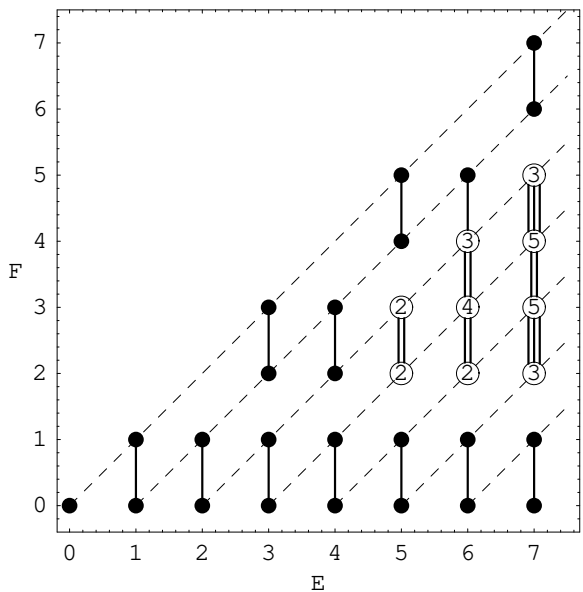}
\end{center}
\vskip-4mm
\caption{Chew--Frautschi plot of the $F=0$ and $F=1$ eigenenergies at weak 't Hooft coupling.}
\label{fig:cf}
\end{figure}

An amusing mathematical digression can be made at this point: our states are, in many ways, what mathematicians would call ``binary necklaces" because they have a cyclic structure and are made out of just two kinds of  ``beads", a bosonic and a fermionic one.
 One immediately finds on the simplest examples that, in general, necklaces of a given (even) length do not split in an equal number of necklaces with even or odd fermions (there are always more bosonic than fermionic necklaces). This would immediately violate SUSY, except that Pauli's principle comes to the rescue by killing a subset of bosonic necklaces. Such a result, which can  obviously be formulated as a purely mathematical  combinatorial problem, appears to be new\footnote{GV wishes to thank Professor  J.C. Yoccoz for this private information.}.

Coming back to our CFP, we may discuss further which linear combinations of the degenerate states of a given $F$ and $E$ form SUSY doublets. The task is facilitated by introducing another conserved
quantity \cite{VW1}:
\eq
C \equiv [\Qd, Q]    \, , ~  [C, H] =0
\eqx
and by noting that  $C^2 = H^2$. Clearly the eigenstates  can be classified according to their
``$C$-parity" i.e. according to whether $C = \pm E$. We may even consider the combination of $C$ and fermion number $C_F = (-1)^F C$.
States with $C=+1~ (-1)$ are clearly those annihilated by $\Qd~ (Q)$. Most of the states turn out to have
$C_F= -1$ (all of them in the $F=0,1$ sectors), but there also are some ``unnatural" ones  with $C_F = +1$, which are very important for the full matching of bosons and fermions discussed above.
While the analysis of generic values of $F$ needs a more systematic --and probably computer-based-- approach, we have been able to deal by brute force with the $F=2,3$ sectors \cite{VW2}, confirming
all the results discussed above.
There is, however, another surprise:
As $\lambda \rightarrow 1$, all states again go to $E=0$;  at  $\lambda > 1$,
 {\it two}  states with $F=2$ stay at $E=0$, while all other states move up  to positive $E$.
 The question of whether new $E=0$ states keep popping up at large $\lambda$ in even higher-$F$ sectors  is currently  being explored.

\section{Outlook}

We would like to conclude this paper by  outlining a possible programme for connecting its  two parts
in a  reasonably near future.

A  first step would consist in generalizing our SQM model to the case in which bosons and fermions belong to
different representations of the symmetry group $U(N)$. In the case of bosons in the adjoint
representation and fermions in the two-index (anti)symmetric  representation (plus its complex conjugate),
it would be interesting to check whether (at least parts of) the bosonic spectra of the latter theories coincide with those
of the supersymmetric case in the planar limit.

Next, we would like to extend our planar calculus to systems with rotational symmetry, such as those obtained
by  dimensional reduction of (supersymmetric) Yang--Mills theories in $D=3+1$ and $D=9+1$ dimensions.
This basically requires generalization of the planar rules for more than one species of bosons and fermions.
A good step in this direction is being made by studying our supersymmetric  quantum mechanical model
with an arbitrary number of fermions \cite{OVW}.

Should these first two steps  succeed, the next one would be to extend the whole approach to the planar
Hamiltonians
of full fledged quantum field theories, with or without supersymmetry, and in particular
to the  orientifold large-$N$ limit of QCD described in the first part of this paper.

Finally,  comparing the $N = \infty$ results with those that can be obtained at $N=3$ by the method
of Ref.  \cite{CW}, would give an estimate of how good the orientifold $1/N$ expansion is at  $N=3$
 and  of the extent to which  supersymmetric
gauge theories can make accurate predictions for actual QCD.

\section*{Acknowledgements}
GV and JW wish to  thank A. Armoni, P. van Baal, L. Giusti, V. Kazakov, G. Marchesini, P. Menotti,
E. Rabinovici, A. Schwimmer, G. Shore and K. Zalewski for useful discussions and/or correspondence on several of the topics discussed here.
Most of all, we would like to thank the organizers of this meeting
for providing an opportunity to discuss so many interesting subjects with Adriano's numerous friends and colleagues in a very pleasant setting.
This work is partially supported
by the grant of the Polish Ministry of Science and Education P03B 024 27 (2004--2007).

\end{document}